\documentclass[%
 reprint,
groupedaddress,
 amsmath,amssymb,
 aps,
prx,
]{revtex4-1}

\usepackage{graphicx}
\usepackage{dcolumn}
\usepackage{bm}

\begin{document}

\newcommand{\be}{\begin{equation}}
\newcommand{\ee}[1]{\label{#1}\end{equation}}
\newcommand{\bem}{\begin{eqnarray}}
\newcommand{\eem}[1]{\label{#1}\end{eqnarray}}
\newcommand{\eq}[1]{Eq.~(\ref{#1})}
\newcommand{\Eq}[1]{Equation~(\ref{#1})}
\newcommand{\vp}[2]{[\mathbf{#1} \times \mathbf{#2}]}


\title{ Interplay of spin and mass  superfluidity in antiferromagnetic  spin-1 BEC \newline and bicirculation vortices}

\author{E.  B. Sonin}
\email[]{sonin@cc.huji.ac.il}
 \affiliation{Racah Institute of Physics, Hebrew University of
Jerusalem, Givat Ram, Jerusalem 91904, Israel}

\date{\today}

\begin{abstract}
The paper investigates the coexistence and interplay of spin and mass superfluidity in the antiferromagnetic spin-1 BEC. The hydrodynamical theory describes the spin degree of freedom by the equations similar to the Landau--Lifshitz--Gilbert theory for bipartite antiferromagnetic insulator. The variables in the spin space are two subspins with  absolute value $\hbar/2$, which play the role of two sublattice spins in the antiferromagnetic insulators. 

As well as in bipartite antiferromagnetic insulators,  in the antiferromagnetic spin-1 BEC there are two spin-wave modes, one is a gapless Goldstone mode, another is gapped.  The Landau criterion shows that in limit of small total spin (two subspins are nearly antiparallel) instability of supercurrents starts from the gapped mode. In the opposite limit of large total spin (two subspins are nearly parallel) the gapless modes become unstable earlier than the gapped one. Mass and spin supercurrents decay via phase slips, when vortices cross streamlines of supercurrent. The vortices participating in phase slips are nonsingular bicirculation vortices. They are characterized by two topological charges, which are winding numbers  describing circulations of two angles around the vortex axis. The winding numbers can be half-integer. A particular example of a half-integer vortex is a half-quantum vortex with the superfluid velocity circulation $h/2m$. But  the superfluid velocity circulation is not a topological charge, and in general the quantum of this circulation can be continuously tuned from 0 to $h/2m$.
\end{abstract}

\maketitle



\section{Introduction} \label{Intr} 

Spin superfluidity in magnetically ordered systems is discussed a number of decades \cite{ES-78b,ES-82,Bun,Adv,BunV,Tserk,Halp,Mac,Pokr,Son17,Hoefer,BratAF}. 
 The phenomenon is based on the analogy of special cases of the Landau--Lifshitz--Gilbert (LLG) theory in magnetism and superfluid hydrodynamics. While in a superfluid mass (charge in superconductors) can be transported by a current proportional to the gradient of the phase of the macroscopical wave function, in a magnetically ordered medium there are spin currents, which are proportional to the gradient of the spin phase. The latter is defined as the angle of rotation around some axis in the spin space. Strictly speaking this analogy is complete only if this axis is a symmetry axis in the spin space. Then according to Noether's theorem the spin component along this axis is conserved. But possible violation of the spin conservation law  usually is rather weak because it is related with relativistically small (inversely proportional to the speed of light) processes of spin-orbit interaction. In fact, the LLG theory itself is based on the assumption of weak spin-orbit interaction \cite{LLstPh2}.

 The analogy of the LLG theory with the theory of superfluidity suggests a new useful language for description of phenomena in magnetism, but not a new phenomenon. During the whole period of spin superfluidity investigations and up to now there have been disputes about definition what is spin superfluidity. There is a school of thinking that the existence of spin current proportional to the spin phase (rotation angle) means spin superfluidity \cite{BunV}. There is no law in the book, which forbids the use of this definition.  But then spin superfluidity becomes a trivial ubiquitous  phenomenon existing in any magnetically ordered medium. A spin current proportional to the spin phase emerges in any domain wall and in any spin wave. Under this broad definition of spin superfluidity spin superfluidity was experimentally detected beyond reasonable doubt in old experiments of the middle of the 20th century detecting domain walls and spin waves.

We prefer define the term superfluidity in its original meaning known from the times of Kamerlingh Onnes and Kapitza: transport of some physical quantity (mass, charge, or spin) over macroscopical distances without essential dissipation. This requires a constant or slowly varying phase gradient at macroscopical scale with the total phase variation along the macroscopic sample equal to $2\pi$ multiplied
by a very large number. In examples of domain walls and spin waves this definitely does not take place. Gradients oscillate in space or time, or in both. The total phase variation is on the order of $\pi$ or much less. Currents transport spin on distances not more than the domain wall thickness, or the spin wavelength. Although such currents are also sometimes called supercurrents, we use the term supercurrent only in the case of macroscopical supercurrent persistent at large spatial and temporal scales. 

The possibility of supercurrents is conditioned by special  topology of the magnetic order parameter space (vacuum manifold).  Namely, this space must have topology of circumference on the plane. In magnetically ordered systems this requires the presence of easy-plane uniaxial anisotropy.  It is possible also in non-equilibrium coherent precession states, when spin pumping supports coherent spin precession with fixed spin component along the magnetic field (the axis $z$). Such non-equilibrium coherent precession states were considered as manifestation of magnon BEC \cite{Safon,Feld}. 
These states  were experimentally investigated in the $B$ phase of superfluid $^3$He \cite{Bun} and in YIG films \cite{Dem6}.  Resemblance and distinction  of coherent precession states with BEC and lasers was discussed in Ref.~\onlinecite{Adv}.

For assessment of possibility of observation of long-distance spin transport  by spin supercurrents one should consider  the Landau criterion, which checks stability of supercurrent states with respect to weak excitations  of all collective modes. Although the Landau criterion points out a  threshold for the current state instability, it tells nothing about how the instability develops. The decay of the supercurrent is possible only via phase slips. In a phase slip event a vortex crosses current streamlines decreasing the phase difference along streamlines. Below some critical value of supercurrent phase slips are suppressed by energetic barriers. The critical value of the supercurrent at which barriers vanish is of the same order as those estimated from the Landau criterion. This leads to a conclusion that the instability predicted by the Landau criterion is a precursor of the avalanche of phase slips not suppressed by any activation barrier.

Recently investigations of spin superfluidity were extended to spin-1 BEC, where spin and mass superfluidity coexist and interplay. Investigations focused on the ferromagnetic BEC \cite{LamSpin,Duine,Sp1}.   The present paper extends the analysis on all states of spin-1 BEC, either ferro- or antiferromagnetic. The interplay of spin and mass superfluidity is responsible for a number of new nontrivial features.

The first step of the analysis was reformulation of hydrodynamics of spin-1 BEC presenting it in the form more suitable for the goals of this paper.  It was already known that hydrodynamics of ferromagnetic spin-1 BEC is described by equations of spin motion  similar to those in the LLG theory in magnetism but taking into account the possibility of superfluid motion as a whole \cite{LamSpin,Duine,Sp1}.  This allowed to use some results known from investigations of spin superfluidity in magnetically ordered solids. In the present paper we demonstrate that the hydrodynamics of the antiferromagnetic spin-1 BEC is similar to the LLG theory for a bipartite antiferromagnet with two sublattices each of which is characterized by a vector of magnetization (spin). Despite translational invariance is not broken and there are no sublattices in the spin-1 BEC, one can introduce two spins of absolute value $\hbar/2$, which  can vary their direction in space and time, but not their absolute values similarly to two  sublattice magnetizations in the LLG theory for a bipartite antiferromagnet. We shall call them subspins. Thus,   results of the recent analysis of spin superfluidity in solid antiferromagnets with localized spins \cite{Son19} become relevant for antiferromagnetic spin-1 BEC. In particular, like in solid antiferromagnets, in the antiferromagnetic spin-1 BEC there are two spin-wave modes: one is a Goldstone gapless mode similar to that in the ferromagnetic BEC and another has a gap depending on the magnetic field. At weak magnetic fields (the Zeeman energy is less than the spin-dependent interaction energy) the Landau critical values are reached in the gapped mode earlier than in the gapless one. At strong magnetic fields close to the field at which spin polarization is completely saturated along the magnetic field, the situation is opposite: the gapless mode becomes unstable earlier than the gapped one. 

 While in scalar superfluids the Landau critical velocities for mass currents are scaled by the sound velocity and in magnetically ordered media with localized spins critical velocities for spin currents are scaled by the spin-wave velocity, in the spin-1 BEC, where spin and mass superfluidity coexist, critical velocities for both currents are determined by the lesser from the sound and spin-wave velocity. Usually this is the spin-wave velocity, and the analysis of the paper focuses on the spin degree of freedom. Another remarkable outcome of  the interplay of mass and spin superfluidity  is properties of vortices participating in phase slips. In the multicomponent  spin-1 BEC vortices are determined not by one but by two winding numbers (topological charges). The two charges are related with two gauge invariances: with respect to the global phase of the wave function and with respect to the spin phase.  We call these vortices bicirculation vortices.

 In the spin-1 BEC, like in other multicomponent superfluids,  the circulation of superfluid velocity is not a topological charge anymore, and the superfluid velocity is not curl-free.
 In single-component scalar superfluids, where the velocity circulation is a topological charge, the velocity field around the vortex is singular and diverges as $1/r$, when   
the distance $r$ from the vortex axis goes to 0. The divergence in the energy can be avoided only if the superfluid density vanishes at the vortex axis. But in the spin-1 BEC the singularity $1/r$ can be compensated without suppression of the superfluid density in the vortex core by a proper choice of the ratio between two topological charges (winding numbers). Such vortices  are called nonsingular or continuous \cite{Sal85}. Normally the Landau critical gradients and the critical gradients for the instability with respect to phase slip are of the order of the inverse core radius.  Therefore, the instability with respect to phase slips starts earlier for nonsingular vortices because of their larger core radius compared to singular vortices. 
Existence of vortices with different ratios of two winding numbers makes  the decay of supercurrents more complicated. There are vortices, which are  effective for relaxation of mass supercurrents, and  which are  effective for relaxation of spin supercurrents. For complete relaxation of all supercurrents to the ground state at least  two different types of vortices must participate in phase slips.

In spin-1 BEC  winding numbers  of vortices  can be not only integer, but also half-integer. The known example of half-integer vortices is the half-quantum vortex, which attracted a lot of attention in the literature \cite{Vol1/2,UedaR}.  However, in the spin-1 BEC the ``quantum'' of velocity circulation can be equal not only to the fundamental quantum $h/m$ or its half, but in fact can be continuously tuned  by a magnetic field, or by the intensity of spin pumping, which supports the non-equilibrium coherent precession state with fixed $z$-component of spin.

\section{Hydrodynamics from the Gross--Pitaevskii theory of spin-1 BEC}

The wave function  of bosons with spin 1  is a 3D vector in the spin space, In the Cartesian basis \cite{Ho98,Ohmi1,UedaR} the wave function vector is
\be
\bm \psi   =\left(\begin{array}{c}\psi_x \\ \psi_y\\ \psi_z\end{array} \right),
     \ee{psi}
where 
\be
\psi_x={\psi_+ -\psi_-\over \sqrt{2}},~~\psi_y={i(\psi_+ +\psi_-)\over \sqrt{2}},~~\psi_z=-\psi_0,  
    \ee{}
and  $\psi_\pm,\psi_0$  are coefficients of the expansion of the wave function in eigenfunctions of the spin projection on the quantization axis (the axis $z$) with eigenvalues $\pm1,0$.   

The Gross--Pitaevskii equation for the wave-function vector $\bm \psi$,  
\bem
i\hbar {\partial \bm \psi \over \partial t}={\delta  {\cal H} \over \delta \bm \psi^*},
    \eem{EqHam} 
is obtained by variation of the Lagrangian 
\be
{\cal L}= {i\hbar\over 2} \left(\bm \psi ^* {\partial \bm \psi \over \partial t}-\bm \psi  {\partial \bm \psi^* \over \partial t}\right)-{\cal H} (\bm \psi ,\bm \psi ^*)
    \ee{}
 with respect to $\bm \psi^*$. The complex-conjugate equation follows from variation with respect to $\bm \psi$. 
Here 
\be
{\delta  {\cal H} \over \delta \bm \psi^*}={\partial {\cal H} \over \partial \bm \psi^*}-\nabla_i {\partial  {\cal H} \over \nabla_i \partial \bm \psi^*}
   \ee{}
is a functional derivative of the Hamiltonian
\bem
{\cal H} ={\hbar^2 \over 2m} \nabla_i\bm \psi^* \nabla_i\bm \psi+{V|\bm \psi|^4\over 2}+{V_s(|\bm \psi|^4-|\bm \psi^2|^2)\over 2}
\nonumber \\
-\gamma \bm H\cdot  \bm S |\bm \psi |^2,
      \eem{Ham3}
where $\gamma$ is the gyromagnetic ratio, $\bm H$ is the magnetic field, and  $V$ and $V_s$ are amplitudes of spin-independent and spin-dependent interaction of bosons respectively, Spin-1 BEC is ferromagnetic if $V_s$ is negative and antiferromagnetic if $V_s$ is positive. The complex vector $\bm \psi$ determines the  particle density  $n =|\bm \psi|^2$    of  bosons with  spin per particle
\be
\bm S=-{i\hbar[\bm \psi^* \times \bm \psi]\over |\bm \psi|^2},
        \ee{}
and  the superfluid velocity
\bem 
v_i =-{i\hbar \over 2m|\bm \psi |^2}(\bm \psi ^* \nabla_i \bm \psi -\bm \psi  \nabla_i \bm \psi ^*).
   \eem{}      
The equation of the spin balance is
\be
{\partial (n  S_i)\over \partial t} +\nabla_k J_{ik} = G_i,
       \ee{}
where the tensor
\be
J_{ik} = \epsilon_{ist} \left( \psi^*_s  {\partial  {\cal H}\over    \partial   \nabla_k \psi_t }+\psi_s  {\partial  {\cal H}\over    \partial   \nabla_k \psi^*_t }\right)
     \ee{spCur}
is the current of the  $i$th spin component along the axis $k$, and 
\bem
G_i =  \epsilon_{ist} \left( \psi^*_s  {\partial  {\cal H}\over    \partial    \psi_t }+\nabla_k\psi^*_s  {\partial  {\cal H}\over    \partial   \nabla_k \psi_t }
\right.\nonumber \\  \left.
+ \psi _s  {\partial  {\cal H}\over    \partial    \psi^*_t }+\nabla_k\psi_s  {\partial  {\cal H}\over    \partial   \nabla_k \psi^*_t }\right)
    \eem{}
is the torque on the  $i$th spin component, which vanishes if the Hamiltonian is spherically symmetric in the spin space.

Now we perform the generalized Madelung transformation presenting the wave function vector as 
\be
\bm\psi=\psi_0 e^{i\Phi}\left(\cos {\lambda\over 2} \bm d +i\sin {\lambda\over 2}  \bm f\right),
          \ee{ddn}
where the real scalar  $\psi_0=\sqrt{n}$, the two real unit mutually orthogonal vectors  $ \bm d$ and $ \bm f$, and the two phase (angle) variables $\Phi$ and $\lambda$ are 6 parameters fully determining the complex 3D vector $\bm \psi$. In new hydrodynamical variables the spin is
\be
\bm S=\hbar\sin \lambda [\bm d \times \bm f],
        \ee{}
and the superfluid velocity is 
\bem 
v_i = {\hbar\over m}\left[\nabla_i \Phi + {\sin \lambda \over 2}(\bm d\nabla_i \bm f-\bm f\nabla_i \bm d)\right].
   \eem{}

The two unit vectors $\bm d$ and $\bm f$ together with the third unit vector,
\be
\bm s={\bm S\over S}=[\bm d \times \bm f],
     \ee{}
fully determine the quantum state  for the ferromagnetic spin-1 BEC, when $\lambda =\pi/2$. States with $0 \leq  \lambda <\pi/2$ are  antiferromagnetic states with the absolute value $S$ of the total spin less than its maximal value $\hbar$ in the ferromagnetic state.  The pure antiferromagnetic state with zero total spin ($\lambda=0$) is called polar phase \cite{UedaR}. 

The Hamiltonian (\ref{Ham3}) transforms to 
\be 
{\cal H}= {mn v^2\over 2} +{\cal H}_0, 
   \ee{}
where 
\bem 
{\cal H}_0={n\hbar^2\over 2m}\left[\cos^2{\lambda \over 2}  \nabla \bm d^2+\sin^2{\lambda \over 2}  \nabla \bm f^2 -\sin^2 \lambda (\bm f \nabla \bm d)^2
\right. \nonumber \\ \left.
+{\nabla \lambda^2\over 4}\right]+{Vn^2\over 2} +{V_sn^2\sin^2\lambda \over 2}-\gamma n\hbar\sin \lambda \bm H\cdot[\bm d\times \bm f]~~~~~
   \eem{}
is the Hamiltonian in the coordinate frame moving with the superfluid velocity $\bm v$.

The dynamical equations for our hydrodynamical variables follow from the  nonlinear Schr\"odinger equation (\ref{EqHam}):
\bem
\hbar[\dot \Phi  +(\bm v \cdot \bm \nabla)\Phi]= -{\delta {\cal H}_0 \over \delta n},
\nonumber \\
\dot n={1\over \hbar}{\delta {\cal H}\over \delta  \Phi} = -{1\over \hbar}\nabla \cdot {\partial {\cal H}\over\partial \nabla \Phi}=-\nabla \cdot (n \bm v),
     \eem{mdf}  
  \bem
n\hbar \cos  \lambda[\dot  \lambda  +(\bm v \cdot \bm \nabla)  \lambda] = -\left(\bm f {\delta {\cal H}_0\over \delta \bm d} - \bm d  {\delta {\cal H}_0\over \delta \bm f} 
\right),
   \eem{lam}
   \bem  
n\hbar [\dot  {\bm d}  +(\bm v \cdot \bm \nabla) \bm d]  ={\bm f\over \cos \lambda}  {\delta {\cal H}_0\over \delta \lambda}+{\bm  s\over \sin \lambda}  \left( \bm s {\delta {\cal H}\over \delta \bm f} \right),
   \eem{d}
\be 
n\hbar [\dot  {\bm f}  +(\bm v \cdot \bm \nabla) \bm f]  =-{  \bm d\over \cos \lambda }  {\delta {\cal H}_0\over \delta \lambda}-{  \bm s\over \sin \lambda} \left(  \bm s {\delta {\cal H}\over \delta \bm d}\right).
  \ee{dn} 
The continuity equation [the second line of \eq{mdf}] takes into account  that    because of   gauge invariance the Hamiltonian depends on the gradient of $\Phi$ but not  on the phase $\Phi$ itself.   
The superfluid velocity is not curl-free and the generalized Mermin--Ho relation is 
\bem
\bm \nabla \times \bm v={\hbar\over 2m}\{ \cos \lambda  [\bm \nabla \lambda \times  ( d_i \bm \nabla  f_i-f_i \bm \nabla d_i) ]
\nonumber \\
+\sin \lambda \epsilon_{ijk} s_i [\bm \nabla s_j \times \bm \nabla s_k]\}.
                                              \eem{9.2}
In the ferromagnetic state ($\lambda =\pi/2$) this reduces to the original Mermin--Ho relation \cite{Mer}.

We want to demonstrate the analogy of the spin-1 BEC hydrodynamics with LLG theory for bipartite antiferromagnet. The vector 
$ \bm L \, =\hbar \cos \lambda \bm d$ is an analogue of the antiferromagnetic vector (staggering magnetization) in the LLG theory of a bipartite antiferromagnet. Continuing  this analogy,   we may introduce the spins $\bm S_1$ and  $\bm S_2$ similar to spins of two sublattices of a bipartite antiferromagnet, which determine the antiferromagnetic vector $\bm L=\bm S_1-\bm S_2$ and the total spin  $\bm S=\bm S_1+\bm S_2$. The vectors $\bm L$ and $\bm S$ are orthogonal one to another,  and   the absolute values of  vectors $\bm S_1$ and  $\bm S_2$ are equal to $S_0=\hbar /2$ and  do not vary in space and time, similarly to sublattice magnetizations  in a bipartite antiferromagnet.

The spins  $\bm S_1$ and  $\bm S_2$ may replace $\lambda$, $\bm d$, and   $\bm f$ as hydrodynamical variables. Then the canonical equations for the spin degree of freedom become  
\bem
n[ \dot {\bm S}_i+(\bm v \cdot \bm \nabla) \bm S_i]= -\left[\bm S_i \times {\delta {\cal H}_0\over \delta \bm S_i }\right],
  \eem{LL}
where $i=1,2$. 
Apart from the term $\propto \bm  v$ in the left-hand side taking into account the superfluid motion as a whole, \eq{LL} is exactly the LLG equations for a bipartite antiferromagnet.  The Hamiltonian ${\cal H}_0$ in \eq{LL} directly follows from the Hamiltonian in the Gross--Pitaevskii theory:
\begin{widetext}
\bem
{\cal H}_0={n\over 2m}\left[{(\nabla \bm S_1+\nabla \bm S_2)^2\over 2(1+\cos \lambda)}+(\nabla \bm S_1-\nabla \bm S_2)^2
+\left(1-{2\cos^2 \lambda\over 1+\cos \lambda }  - 4\sin^2 \lambda  \right){\nabla (\bm S_1\cdot \bm S_2)^2\over \sin^2 2\lambda}
+ {( \bm S_1 \nabla \bm S_2-\bm S_2 \nabla \bm S_1)^2\over  (1+\cos \lambda)\cos \lambda}   \right]
\nonumber \\
+{V_sn ^2(\bm S_1+ \bm S_2)^2\over 2\hbar^2} -n \gamma \bm H \cdot(\bm S_1+ \bm S_2).~
   \eem{H0}
\end{widetext}
 In the LLG theory for localized spins they usually use the Hamiltonian, which  is a general quadratic form of gradients $\nabla \bm S_1$ and $\nabla \bm S_2$ with constant coefficients. In \eq{H0} the coefficients depend on the angle $\lambda$, which depends on $ \bm S_1$ and $\bm S_2$:
\be
\cos 2\lambda =-{(\bm S_1\cdot \bm S_2)\over S_0^2}=-{4(\bm S_1\cdot \bm S_2)\over \hbar^2}.
     \ee{}
 In polar angles determining directions of $\bm S_1$ and  $\bm S_2$,
\bem
S_{ix}=S_0 \cos \theta_i \cos\varphi_i,~~S_{iy}=S_0 \cos \theta_i \sin\varphi_i,
\nonumber \\
S_{iz}=S_0\sin\theta_i,
   \eem{thePhi}
\eq{LL} transforms to 
\bem
 \cos \theta_i[\dot \theta_i+(\bm v \cdot \bm \nabla) \theta_i]=-{1\over nS_0}{\delta {\cal H}_0\over \delta \varphi_i} , 
 \nonumber \\ \
 \cos \theta_i[\dot\varphi_i  +(\bm v \cdot \bm \nabla) \varphi_i ]={1\over nS_0}{\delta {\cal H}_0\over \delta \theta_i}.
     \eem{tipi}

Of course, there are no lattices or sublattices in the spin-1 BEC. But this strong analogy with an antiferromagnet with two spin sublattices means that the LLG theory remains valid even if spins are delocalized and sublattices melt down. In a bipartite antiferromagnet the angle $\lambda$ is a canting angle measuring deviation of sublattice spins from the strictly antiparallel orientation in a pure antiferromagnetic state with zero total spin. We shall call spin vectors $\bm S_1$ and $\bm S_2$ subspins.

In cold atoms BEC the  spin-independent interaction $\propto V$ is much stronger than the spin-dependent one  $\propto V_s$.   According to \citet{Ho98},  the ratio $|V_s|/V$ is 0.04 for $^{23}$Na and 0.01 for $^{97}$Ru. Correspondingly, the ratio of the spin-wave velocity to the sound velocity proportional to $\sqrt{|V_s|/V}$ is also small.  This allows in the further analysis to ignore \eq{mdf} describing motion of the superfluid as a whole and to assume that the superfluid is incompressible. This does not rule out possibility of superfluid mass currents with $\bm v\neq 0$,  but the velocity must be divergence-feee:  $\bm \nabla \cdot \bm v=0$.  Stability of  mass currents is also can be investigated considering only the spin degree of freedom and ignoring the mechanical degree of freedom. At  was shown in Ref.~\onlinecite{Son19}   for ferromagnetic spin-1 BEC and will be shown below for antiferromagnetic spin-1 BEC, instability starts in the softest mode, spin mode in our case.
So our further analysis focuses on Eqs.~(\ref{lam})--(\ref{dn}) with the Hamiltonian \eq{H0}. 

\section{Collective modes and the Landau criterion}

For the further analysis it is convenient to transform the angle variables in \eq{thePhi} to angles
\bem
\theta_0={\pi +\theta_1-\theta_2\over 2},~~  \theta ={\theta_1+\theta_2-\pi\over 2},
\nonumber \\
\varphi_0={\varphi_1+\varphi_2\over 2},~~\varphi={\varphi_1-\varphi_2\over 2},
    \eem{angleU}
which have already been used in the analysis of spin dynamics in antiferromagnetic  insulators \cite{Son19}. A benefit of these variables is that the dynamical equations are reduced to decoupled equations for two noninteracting modes. In these variables the equations of motion \eq{tipi} transform to
\bem
(\cos 2\theta_0+ \cos 2\theta)[\dot \theta_0+(\bm v \cdot \bm \nabla) \theta_0]
\nonumber \\
=-{1\over nS_0}\left(\cos \theta_0 \cos \theta{\delta {\cal H}_0\over \delta \varphi_0} +\sin \theta_0 \sin \theta{\delta {\cal H}_0\over \delta \varphi}\right),
\nonumber \\
(\cos 2\theta_0+ \cos 2\theta)[\dot \varphi_0+(\bm v \cdot \bm \nabla) \varphi_0]
\nonumber \\
={1\over nS_0}\left(\cos \theta_0 \cos \theta{\delta {\cal H}_0\over \delta \theta_0} -\sin \theta_0 \sin \theta{\delta {\cal H}_0\over \delta \theta}\right),
    \eem{}
\bem
(\cos 2\theta_0+ \cos 2\theta)[\dot \theta+(\bm v \cdot \bm \nabla) \theta]
\nonumber \\
={1\over nS_0}\left(\cos \theta_0 \cos \theta{\delta {\cal H}_0\over \delta \varphi} +\sin \theta_0 \sin \theta{\delta {\cal H}_0\over \delta \varphi_0}\right),
\nonumber \\
(\cos 2\theta_0+ \cos 2\theta)[\dot \varphi+(\bm v \cdot \bm \nabla) \varphi]
\nonumber \\
=-{1\over nS_0}\left(\cos \theta_0 \cos \theta{\delta {\cal H}_0\over \delta \theta} -\sin \theta_0 \sin \theta{\delta {\cal H}_0\over \delta \theta_0}\right).
    \eem{}

 \begin{figure}[t]
\includegraphics[width=.3\textwidth]{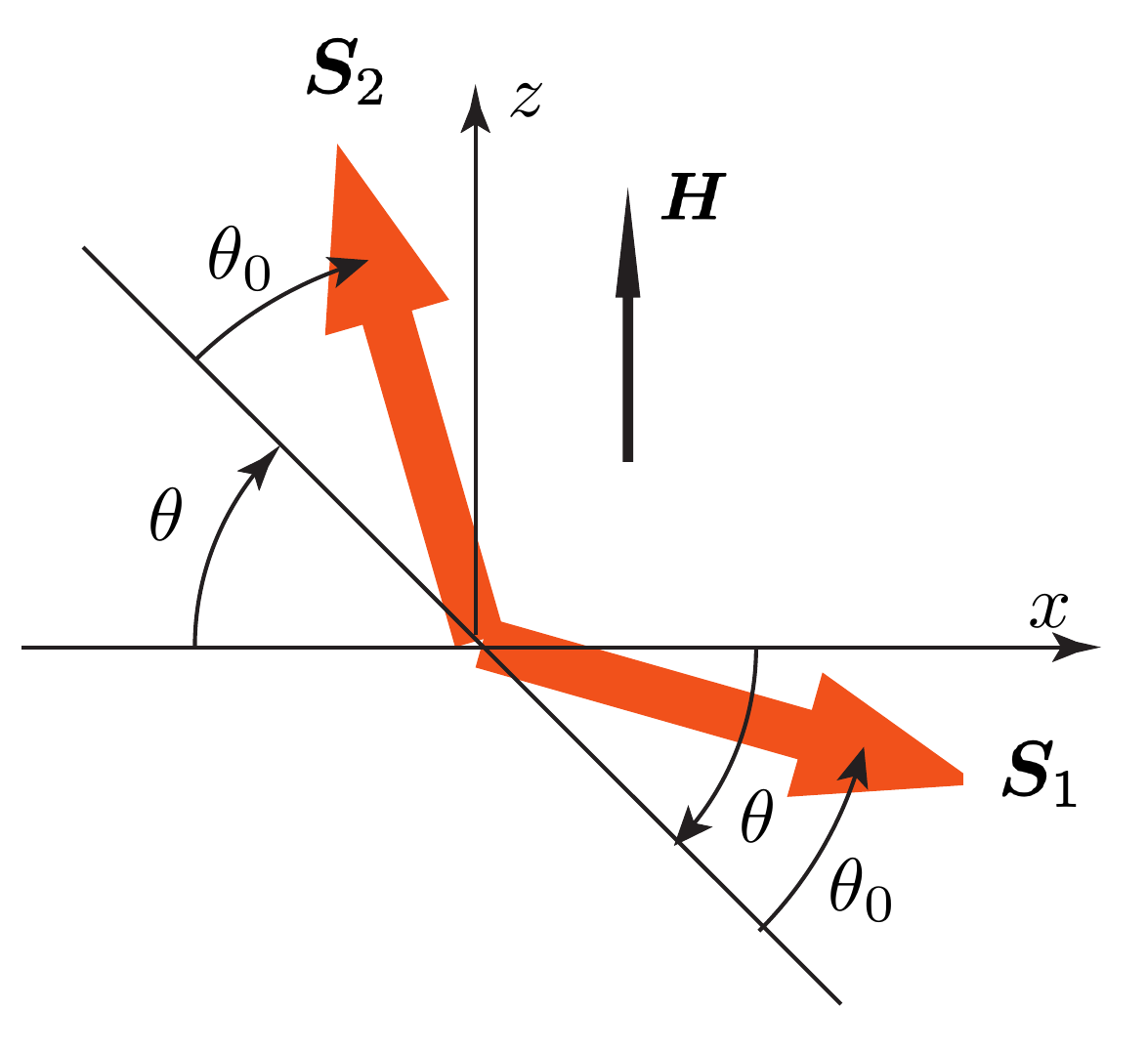}
\caption[]{ Angle variables $\theta$ and $\theta_0$  for the case when the both subspins are in the plane $xz$ ($\varphi_0 =\varphi=0$).}
\label{f1}
\end{figure}

The canting angle $\lambda$ is given by 
\be
\cos 2\lambda={\cos 2\theta_0 (1 +\cos 2\varphi)-\cos 2\theta(1- \cos 2\varphi)\over 2}.
   \ee{}
The meaning of the angles $\theta_0$ and $\theta$ for the simple case $\varphi=\varphi_0=0$ is illustrated in Fig.~\ref{f1}.

The Hamiltonian ${\cal H}_0$ in the angle variables \eq{angleU} is gauge-invariant with respect to phases $\Phi$ and $\varphi_0$. The general expression for it  rather clumsy, and further we shall present the Hamiltonian only for particular cases. We consider  states with $\varphi=0$ when the canting angle $\lambda$ coincides with the angle $\theta_0$. Then the Hamiltonian is 
\bem
{\cal H}={nm v^2\over 2}+{n\hbar^2\over m}\left[{\nabla \theta_0^2\over 8}+\nabla \theta^2 {1+\cos \theta_0\over 4}
\right. \nonumber \\ \left.
+\nabla \varphi_0^2 \left({1-\cos \theta_0\over 4}\sin^2\theta+ {\cos^2\theta_0\cos^2\theta\over 2}\right)\right] +E_m,
\nonumber \\
   \eem{AFv}
where the superfluid velocity is
\be
\bm v={\hbar\over m}(\bm \nabla\Phi -\sin\theta_0\cos\theta \bm\nabla \varphi_0),
     \ee{vth0th}
and the energy $E_m$ includes the spin-dependent interaction $\propto V_s$,  and the Zeeman energy:
\be
E_m={V_sn^2 \over 2}  \sin^ 2\theta_0 -\hbar n \gamma H   \sin\theta_0\cos\theta.
   \ee{AFvFM}

The current of the  spin $z$-component is given by \eq{spCur} for $i=z$ and can be written as
\be
J_{zk} = n S_z  v_k +J_k,
     \ee{}
where the first term on the right-hand side is an advection term, which is related with spin transport by motion of the whole superfluid and has nothing to do with spin superfluidity. We determine the spin supercurrent  in the coordinate frame moving together with the superfluid. In the angle variables the spin supercurrent is a vector
\be
\bm J =-{n\hbar^2\over m}\left[{\sin^2\theta_0\sin^2 \theta\over 2 (1+\cos \theta_0) }
+\cos^2\theta_0\cos^2 \theta\right]\bm \nabla \varphi_0.
        \ee{}

Considering the collective mode connected with oscillations of $\theta_0$ and $\varphi_0$, one can further simplify the Hamiltonian assuming that $\theta=0$ in Eqs.~(\ref{AFv})--(\ref{AFvFM}). Then the equations of motion are
\bem
\cos \theta_0[\dot \theta_0+(\bm v \cdot \bm \nabla) \theta_0]=\bm \nabla \cdot  \left({\hbar \cos ^2\theta_0 \bm  \nabla \varphi_0\over m}\right),
\nonumber \\
\dot \varphi_0+(\bm v\cdot \bm \nabla) \varphi_0=-{\hbar \sin\theta_0\nabla \varphi_0^2\over m} -{\hbar \nabla^2 \theta_0\over 4m\cos \theta_0} 
\nonumber \\
+{V_sn \over \hbar}\sin \theta_0-\gamma H.
      \eem{ThPhio}

We address the state with stationary uniform currents characterized by two constant gradients $\bm K_\Phi=\bm \nabla \Phi$ and  $\bm K=\bm  \nabla \varphi_0$.  In this state the spins precess around the  axis $z$ with a constant precession frequency
\be
\dot \varphi_0=-\bm v\cdot  \bm K+\left({V_sn \over \hbar}-{\hbar K^2\over m}\right)\sin \theta_0-\gamma H.
      \ee{}

One may consider the field $\bm H$ not as an external magnetic field, but as a Lagrange multiplier determined from the condition $\delta {\cal H}/\delta \theta_0=0$. Then the precession in the current state is absent in the laboratory coordinate frame.

Next we linearize \eq{ThPhio} with respect to small oscillations $\theta_0'$ and $\varphi_0'$ around the constant values of $\theta_0$ and $\varphi_0$ in the stationary current state:
\bem
\dot \theta_0'+ (\bm w_0\cdot\bm\nabla)\theta_0' ={\hbar \cos \theta_0 \nabla ^2\varphi_0'\over m}
\nonumber \\
\dot \varphi_0'+(\bm w_0\cdot\bm\nabla) \varphi_0'-\bm K\cdot \bm v'=-{\hbar \nabla^2 \theta_0'\over 4m\cos \theta_0}
\nonumber \\
+\left({V_sn \over \hbar }-{\hbar K^2\over m}\right)\cos \theta_0\theta_0'
      \eem{lin}
where
\be
\bm w_0=\bm v +2\sin\theta_0 {\hbar \bm K\over m}={\hbar\over m}(\bm K_\Phi +\sin\theta_0 \bm K)
    \ee{w-0}
is the Doppler velocity $\bm w_0$, which takes into account not only the true Doppler effect (the term $\propto \bm v$), but also the pseudo Doppler effect \cite{Son19} proportional to the gradient $\bm K$ responsible for spin current. 

Although we consider incompressible fluid the superfluid velocity also oscillates but it must be divergence-free:
\be
\bm \nabla \cdot \bm v'={\hbar\over m}\left[ \nabla^2 \Phi' -\sin\theta_0 \nabla^2 \varphi_0' -\cos \theta_0  (\bm K \cdot  \bm \nabla) \theta_0'\right]=0.
   \ee{div} 

 There are plane wave solutions $\theta_0',\varphi_0' \propto e^{i\bm k\cdot \bm r-i\omega t}$ of Eqs.~(\ref{lin}) and (\ref{div}) with the dispersion relation
\be
\omega-\bm w_0\cdot \bm k
=c_sk\sqrt{\left[1-{\xi_0^2 (\bm K\cdot \bm k)^2\over  k^2}\right]\cos^2 \theta_0  + {\xi_0^2  k^2\over 4}},
     \ee{}
where
\be 
c_s=\sqrt{V_sn \over m}
   \ee{}
   is the spin-wave velocity in the ground state without currents and spin polarization ($\theta_0=0$), and 
\be
\xi_0 ={\hbar\over \sqrt {V_s mn}}={\hbar\over mc_s}
      \ee{xiV}
is the coherence length of the spin degree of freedom. This is a gapless Goldstone mode.

The stability condition of the current state (the Landau criterion) requires that the frequency $\omega$ is positive and real for any $\bm k$ . For the sake of simplicity we consider the case when $\bm w_0$ and $\bm K$ are parallel or antiparallel. Instability threshold is minimal  at $\bm k$ parallel to $\bm w_0$.
Then the current state is stable as far as
\be
w_0^2 =\left(\bm v +2\sin\theta_0 {\hbar \bm K\over m}\right)^2<c_s^2 \left(1-\xi_0^2 K^2\right)\cos^2 \theta_0.
  \ee{GoldL}

Another spin-wave mode is connected with oscillations of $\theta$ and $\varphi$. The quadratic in $\theta$ and $\varphi$ correction to  the Hamiltonian of the stationary current state in the  coordinate frame moving with the superfluid is
\begin{widetext}
\bem
{\cal H}_0'={n\hbar^2\over 2m}\left\{{1+\cos\theta_0\over 4}\nabla \theta^2+{\cos^2\theta_0\nabla \varphi^2\over 4(1+\cos\theta_0)} +{\sin 2\theta_0\theta (\bm K\cdot \bm \nabla )\varphi\over 2}
+\left[{V_sn\sin^2 \theta_0 \over \hbar} -{\hbar K^2(1+\cos\theta_0)\over 2m}  \right] {\theta^2\over 2}
\right.\nonumber \\ \left.
+\left[{V_sn \over \hbar} - {\hbar K^2\over 2(1+\cos\theta_0)m} \right]  {\cos^ 2\theta_0 \varphi^2\over 2}  \right\}.
   \eem{} 

Linearized equations of motion for $\theta$ and $\varphi$ after trivial Galilean transformation to the laboratory coordinate frame are
\bem 
\dot \theta+\bm w\cdot \bm  \nabla \theta =-{\cos\theta_0\nabla^2 \varphi\over 2(1+\cos\theta_0)} 
+\left[{V_sn \over \hbar} - {\hbar K^2\over 2(1+\cos\theta_0)m} \right]\cos\theta_0 \varphi,
\nonumber \\ 
 \cos \theta_0(\dot\varphi+\bm w\cdot \bm  \nabla \varphi) ={1+\cos\theta_0\over 2}\nabla^2 \theta
 -\left[{V_sn\sin^2 \theta_0 \over \hbar}  -{\hbar K^2(1+\cos\theta_0)\over 2m}  \right] \theta,
   \eem{gapp}
\end{widetext}
where the Doppler velocity is
\be
\bm w =\bm v +\sin\theta_0 {\hbar \bm K\over m}={\hbar\over m}\bm K_\Phi .
    \ee{w-g}
These equations describe the gapped mode with the spectrum
\be
\omega - \bm w\cdot \bm k={c_s\over \xi_0 }\sqrt{\left[1 -{\xi_0^2 (K^2-k^2)\over 2 } \right]^2-\cos^2\theta_0} .
    \ee{gSp}  
The Landau criterion imposes two inequalities on gradients (velocities) in the current state. The first one, 
\be
 K^2<{2(1-\cos\theta_0)\over \xi_0^2},
    \ee{gapL} 
provides that at no $\bm k$ the frequency has an imaginary part, i.e., the gap in the spectrum is positive. The second inequality,
\be
w^2<{m^2c_s^4\over \hbar^2 k^2}\left\{\left[1 -{\xi_0^2 (K^2-k^2)\over 2 }  \right]^2-\cos^2\theta_0\right\} ,
    \ee{LC2}  
guarantees that the frequency is positive at any $\bm k$. The right-hand side of the inequality has a minimum at
\bem
k^2 ={2 \over \xi_0^2}\sqrt{\left(1 -{\xi_0^2 K^2\over 2 } \right)^2-\cos^2\theta_0}.
   \eem{}
Using this value in \eq{LC2}, the latter becomes 
\be
|w|<c_s\left(\sqrt{\sin^2{\theta_0\over 2} -{\xi_0^2 K^2\over 4}} +\sqrt{\cos ^2{\theta_0\over 2} -{\xi_0^2 K^2\over 4}}\right).
    \ee{}

\begin{widetext}

\begin{figure}[t]
\includegraphics[width=0.9\textwidth]{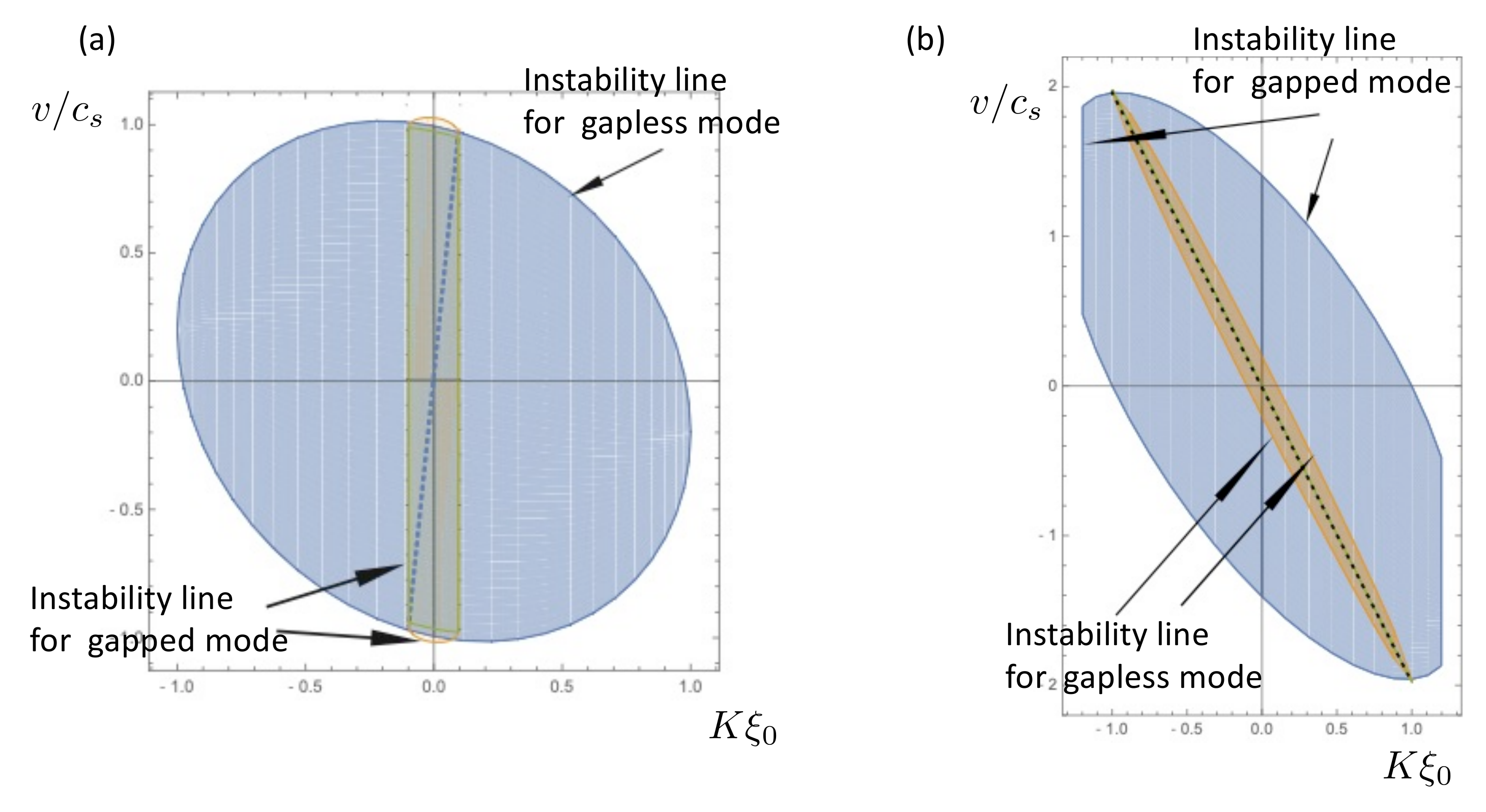}
\caption[]{Stability of two spin-wave modes according to the Landau criterion. (a) Weak spin polarization, $\theta_0 =0.1$. The gapped mode becomes unstable earlier than the gapless mode. The dotted line shows states, to which the current state relaxes after phase slips with antiferromagnetic vortices $(0,\pm 1)$.
 (b) Strong spin polarization, $\tilde\theta_0 =\pi/2-\theta_0 =0.2$. The gapless mode becomes unstable earlier than the gaped mode. The dotted line shows states, to which the current state relaxes after phase slips with ferromagnetic  vortices $(\pm 1/2,\pm 1/2)$.}
\label{fig2}
\end{figure}
\end{widetext}

Figure  \ref{fig2}  shows the stability areas for two modes in the plane of two dimensionless parameters $K\xi_0$ and  $v/c_s$. The current state is stable in the area where the  both modes are stable.  At weak spin polarization (small canting angle $\theta_0$) the gapped mode destabilizes the current state earlier than the gapless one, like in antiferromagnetic insulators \cite{Son19}.  In the opposite limit $\theta_0\to \pi/2$, where the spin polarization is saturated, the gapless modes becomes unstable first.  
For comparison, in  the ferromagnetic spin-1 BEC the only mode is the gapless Goldstone mode, which must be checked with the Landau criterion \cite{Sp1}. 

The Landau criterion checks stability of a current state, but it does not point out what happens when the instability threshold is reached. In scalar superfluids the outcome of instability is disappearance  of potential barriers preventing vortex nucleation and motion of vortices across streamlines (phase slips). In multicomponent superfluids this issue is more complicated because there are various types of vortices. A type of a vortex participating in a phase slip depends on which mode becomes unstable \cite{Sp1}.  We shall return back to the problem of stability of current states after the analysis of vortices possible in the antiferromagnetic spin-1 BEC.

\section{Nonsingular vortices in antiferromagnetic spin-1 BEC}

In scalar (single-component) superfluids only singular vortices are possible, in which the superfluid density must vanish at the vortex axis in order to compensate the $1/r$ divergence in the velocity field.   In multicomponent superfluids it is possible to compensate the divergence at the axis without suppression of the superfluid density in the vortex core.  Since we consider here an incompressible superfluid,   further we shall focus only on phase slips with nonsingular vortices.

Studying nonsingular vortices we can use the Hamiltonian \eq{AFv} derived under assumption that $\varphi=0$. The vortex is characterized  by two topological charges (winding numbers), determined by circulations of the angles $\Phi$ and $\varphi_0$,
\be
N_\Phi= {1\over 2\pi}\oint \bm \nabla\Phi \cdot d\bm l,~~N_\varphi= {1\over 2\pi}\oint \bm \nabla\varphi_0 \cdot d\bm l,
    \ee{}
where integration is along the closed path (loop) around the vortex axis. We shall call these vortices {\it bicirculation vortices} and label them as $(N_\Phi,N_\varphi)$-vortices. 
The superfluid velocity circulation around the path surrounding the vortex at large distances from its axis  is 
\be
\Gamma =\oint \bm v\cdot d\bm l= {h\over m} (N_\Phi - \sin \theta_\infty N_\varphi).
     \ee{}
Here we introduced the angle $\theta_\infty$ equal to the value of $\theta_0$ far from the vortex axis where $\theta=0$ and the gradient-dependent energy is negligible. The angle $\theta_\infty$ depends on the magnetic field and  is determined by minimization of the energy $E_m$ given by  \eq{AFvFM}:
\be
\sin \theta_\infty = {\hbar \gamma H   \cos\theta\over V_sn  }.
   \ee{th-inf}

Single-valuedness of the wave function \eq{ddn} requires that $N_\Phi $ is integer if the unit vectors $\bm d$ and $\bm f$ return back to their original values after going around the path encircling the vortex. But the wave function also remains single-valued if $\bm d$ and $\bm f$ rotate by 180$^\circ$ around the axis normal to both of them, i.e., parallel to the total spin, while the charge $N_\Phi $ is half-integer. The angle of rotation around the spin coincides with the angle $\varphi_0$ only if the spin is strictly parallel or antiparallel to the magnetic field $\bm H$ (axis $z$). In this case the topological charges $N_\varphi$ and $N_\Phi$ can be either both integer, or both half-integer. Correspondingly, we call vortices integer, or half-integer. 

\begin{figure}[b]
\includegraphics[width=0.5\textwidth]{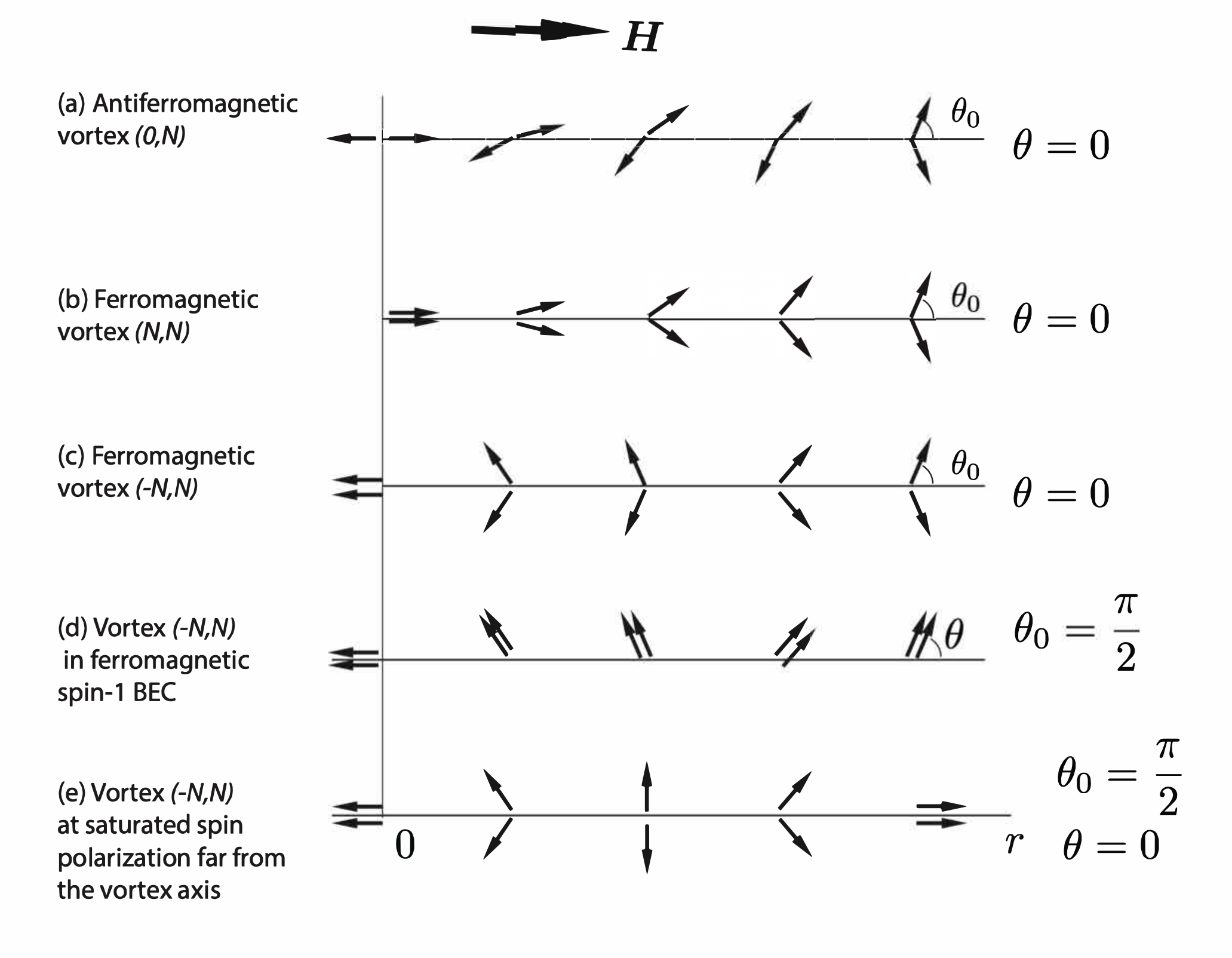}
\caption[]{Variation of two subspins $\bm S_1$ and $\bm S_2$ with the distance $r$ from the vortex axis in skyrmion cores of various types of vortices. The rows (a)--(c) show vortices in the antiferromagnetic spin-1 BEC. The row (d) shows   the vortex  in the ferromagnetic spin-1 BEC. The row (e) shows the vortex $(-N,N)$ at saturated spin polarization far from the vortex axis ($\theta_0\to \theta_\infty =\pi/2 $). This vortex can be half-integer both in the  antiferromagnetic and ferromagnetic spin-1 BEC.}
\label{fig3}
\end{figure}

We shall consider axisymmetric vortices. Then  gradient of the angles  $\Phi$ and  $\varphi_0 $  equal to $\nabla \Phi_v$ and  $\nabla \varphi_v$ respectively have only azimuthal components, 
\be 
\nabla \Phi_v ={N_\Phi[\hat z\times \bm r ]\over r^2},~~\nabla \varphi_v ={N_\varphi[\hat z\times \bm r ]\over r^2},
    \ee{grVor}
 while the angles $\theta_0$ and $\theta$ depend only on the distance $r$ from the vortex axis.  The angle gradients   diverge as $1/ r$. 
 According to the Hamiltonian \eq{AFv} and the expression \eq{vth0th} for the superfluid velocity, this divergence can be compensated only for the two types of vortices:

(i) Any  integer $N_\varphi$, but $N_\Phi=0$. At the vortex axis $\theta_0=0$ and $\theta=\pm {\pi \over 2}$. These are vortices $(0,N)$. 
The structure of the skyrmion core of the vortex is illustrated schematically in Fig.~\ref{fig3}(a) showing variation of  two subspins with the distance $r$ from the vortex axis. 
The vortex can be called antiferromagnetic because the angle $\theta$ showing the direction of the antiferromagnetic vector varies in the vortex core and there is no spin polarization at the vortex axis.  A similar vortex was investigated in the bipartite antiferromagnet in the LLG theory of localized spins \cite{Son19}.

The velocity circulation far from the $(0,N)$ vortex axis is
\be
\Gamma= -{Nh\over m} \sin \theta_\infty. 
    \ee{vcN}
The result \eq{vcN} is remarkable. The velocity circulation is quantized but with the circulation quantum dependent on the magnetic field, since the canting angle $\theta_\infty$ depends on the magnetic field. The quantum varies from zero (i.e., no quantization) to the fundamental quantum $h/m$. 

(ii) Two charges (winding numbers) satisfy the condition $N_\Phi=\pm N_\varphi $, At the axis $\theta_0=\pm {\pi \over 2}$ and $\theta =0$. These are vortices $(\pm N, N)$, which we 
 call ferromagnetic because the spin is fully polarized at the vortex axis. The skyrmion structures of both ferromagnetic vortices are shown in Figs.~\ref{fig3}(b) and (c).
 Ferromagnetic vortices $(\pm N, N)$ have the velocity circulation
 \be
\Gamma= {Nh\over m}(\pm 1-\sin \theta_\infty ).
         \ee{n-n}
The circulation quanta for these vortices can also be tuned by the magnetic field. Since $N$ can be half-integer the circulation quantum varies from 0 to the fundamental quantum $h/m$. The half-quantum vortices \cite{Vol1/2,UedaR} with $\Gamma =\pm h/2m$ are possible in the limit of very weak spin polarization (weak magnetic field), when $\theta_\infty \to 0$.

Investigation of stability of current states with respect to phase slips (the next section) requires knowledge of core radii of various vortices, which can be estimated from the quantitative analysis of vortex cores. Variation of the Hamiltonian  \eq{AFv} 
with respect to  $\theta_0$ and $\theta$  yields two coupled nonlinear Euler--Lagrange equations:
\begin{widetext}
\be
-{1 \over 4r}{d\over dr}\left(r{d \theta_0\over dr}\right) -\left({d \theta\over dr}\right)^2 {\sin \theta_0\over 4}-{N_\Phi N_\varphi \cos \theta_0 \cos\theta\over r^2} 
+{N_\varphi^2\sin \theta_0\sin^2\theta\over 4r^2}
+{(\sin\theta_0 -\sin\theta_\infty \cos\theta) \cos\theta_0\over \xi_0^2}=0,
   \ee{thet0}
\be
-{1 \over 2r}{d\over dr} \left[r{d \theta\over dr} (1+\cos \theta_0)\right]  
+{N_\Phi N_\varphi \sin \theta_0 \sin\theta\over r^2} +{N_\varphi^2\sin 2\theta(1-\cos \theta_0- 2\cos^2\theta_0) \over 4 r^2} 
 +{\sin\theta_\infty \sin\theta_0\sin\theta\over \xi_0^2}=0.
   \ee{tet}
\end{widetext}

Let us start from vortices $(0,N)$  when there is no circulation of the phase $\Phi$.  The $(0,1)$-vortex with minimal winding number $N=1$ is energetically more favorable for phase slips. The natural scale in \eq{tet} is the spin coherence length $\xi_0$, which determined the vortex-core radius $r_c \sim \xi_0$ excepting the limit  of weak spin polarization $\theta_\infty \ll 1$.
In this limit   the size of the vortex core is much larger than the coherence length, as we shall see soon. Then all gradient terms and terms $\propto 1/r^2$ in \eq{thet0} can be neglected and the small $\theta_0$ is determined by a simple expression 
\be
\theta_0(r) =\theta_\infty \cos \theta(r).
    \ee{} 
Inserting it into \eq{tet} transforms it into     
\be
-{1 \over r}{d\over dr} \left(r{d \theta\over dr}\right)
-{\sin 2\theta\over 2 r^2} 
 +{\theta_\infty^2 \sin 2\theta\over 2\xi_0^2}=0.
   \ee{}
The boundary conditions for this equation are $\theta=\pi/2$ at the vortex axis ($r=0$) and $\theta=0$ at $r\to \infty$.
The spatial scale of this equation determines the vortex core radius
\be
r_c \sim {\xi_0 \over \theta_\infty},
      \ee{core01}
 which  essentially exceeds the spin correlation length $\xi_0$. This justifies ignoring of gradient terms in \eq{thet0}.

Switching to vortices $(\pm N, N)$ we also consider the vortices with minimal circulations. These are the $\left(\pm{1\over 2},{1\over 2}\right)$ vortices with half-integer winding numbers. In these vortices  $\theta=0$ everywhere in space, and \eq{thet0} becomes
\be
-{1 \over 4r}{d\over dr}\left(r{d \theta_0\over dr}\right)-{\cos \theta_0 \over 4r^2} 
+{(\sin\theta_0 -\sin\theta_\infty ) \cos\theta_0\over \xi_0^2}=0.
   \ee{}

In the limit  of strong spin polarization when two subspins are nearly parallel, the angle $\tilde\theta_0=\pi/2- \theta$ is small everywhere varying from 0 at the axis of the vortex $\left({1\over 2},{1\over 2}\right)$ to small  $\tilde \theta_\infty$ far from the axis. The equation for $\tilde \theta$,
\be
{1 \over 4r}{d\over dr}\left(r{d\tilde \theta \over dr}\right)-{\tilde \theta \over 4r^2} 
+{(\tilde\theta^2_\infty -\tilde\theta^2 ) \tilde \theta_0\over 2\xi_0^2}=0,
   \ee{}
similar to the equation for the amplitude of the wave function in the Gross--Pitaevskii theory for scalar superfluids. The  vortex core radius is
\be
r_c \sim {\xi_0 \over \tilde\theta_\infty}, 
    \ee{coreF}
    which essentially exceeds the coherence length $\xi_0$ at small $\tilde\theta_\infty$. If the Zeeman energy $\hbar \gamma H$ exceeds  the spin-dependent interaction energy $V_sn$ [see \eq{th-inf}] spin polarization is saturated and the both subspins become parallel to the magnetic field. Then the core radius $r_c$ of the vortex   $\left({1\over 2},{1\over 2}\right)$ given by \eq{coreF} becomes infinite. This vortex  should be ignored because its energy and circulation vanish. However, the half-integer vortex 
 $\left(-{1\over 2},{1\over 2}\right)$  has a finite core radius of the order $\xi_0$, and according to \eq{n-n} its circulation quantum $h/m$ is the same as in the scalar superfluids.     
         
It is interesting to compare the  vortices in antiferromagnetic spin-1 BEC with vortices in ferromagnetic spin-1 BEC, which were considered in Ref.~\onlinecite{Sp1}.  An example of the vortex in ferromagnetic spin-1 BEC  is illustrated in Fig.~\ref{fig3}(d).  In antiferromagnetic spin-1 BEC 
in the vortex core the canting  angle $\theta_0$ varies, while $\theta=0$ everywhere [Fig.~\ref{fig3}(c)]. 
In contrast, in ferromagnetic spin-1 BEC in the vortex core the angle $\theta$  varies, while the canting angle is $\theta_0 =\pi/2$ everywhere. This is because in ferromagnetic spin-1 BEC the interaction constant $V_s$ is negative and the structure with parallel subspins has lesser energy than the structure with antiparallel subspins. In this structure the total spin deviates from the direction of the magnetic field, and this is incompatible with the existence of half-integer vortices.  But at magnetic fields sufficient for  saturated spin polarization along the magnetic field there is a possibility of a half-integer vortex also in ferromagnetic BEC, if the wave function in its core is antiferromagnetic, i.e., $\theta_0 <\pi/2$. The corresponding structure of the skyrmion core is shown in Fig.~\ref{fig3}(e). Thus, a half-integer vortex is possible both in anti- and ferromagnetic phase. But there is a difference in the energy and the size of the core in these two phases.  

In the ferromagnetic spin-1 BEC the spin-dependent interaction energy $V_sn$ is a negative constant, which does not affect the vortex structure. At weak magnetic fields the core radius is scaled not by the coherence length $\xi_0$ given by \eq{xiV} but a much longer length determined by the easy-plane anisotropy energy, which is usually smaller than $|V_s|n$. However, if inside the core the wave function is antiferromagnetic ($\theta_0<\pi/2$) the energy of the core does depend on  $|V_s|n$. As a result, the core radius of the integer vortex (-1,1) is larger than the core radius of the half-integer vortex $\left(-{1\over 2},{1\over 2}\right)$.  This difference vanishes in the limit of very strong magnetic fields when the Zeeman energy essentially exceeds $|V_s|n$. In this limit the core radius of all vortices is determined by the Zeeman energy: $r_c \sim \hbar / \sqrt{ \gamma H S_0 }$.
The integer vortex (-1,1) with  double-quantum velocity circulation $2h/m$ is an analog of the Anderson--Toulouse vortex existing in the A phase of superfluid $^3$He \cite{AnToul,Sal85}.
 
\section{Bicirculation vortices and phase slips}

For the analysis of participation of nonsingular vortices in phase slips one must consider interaction of vortices with mass and spin currents. 
The total energy of the vortex is mostly determined by the area outside the core (the London region) where one must take into account interaction of vortices with mass and spin currents. In the London region the angles $\theta_0$ and $\theta$ are close to their asymptotic values $\theta_\infty$ and 0 respectively. Then only terms quadratic in gradients $\bm \nabla \Phi$ and $\bm  \nabla \varphi_0$ are kept in the Hamiltonian \eq{AFv}.
The phase (angle) gradients in the presence of a vortex and currents are
\be
\bm \nabla \Phi =\bm \nabla \Phi _v +\bm K_\Phi,~~\bm \nabla \varphi_0 =\bm \nabla \varphi _v +\bm K,
    \ee{}
where $\bm \nabla \Phi _v$ and $\bm \nabla \varphi _v $ are given by \eq{grVor}.  Substituting these expressions into the Hamiltonian and integrating over the plane normal to the vortex axis one obtains the energy of the straight vortex per unit length in the presence of currents:
\bem
E_v={\pi n\hbar^2\over  m}(N_\Phi^2 -2\sin \theta_\infty N_\Phi N_\varphi + N_\varphi ^2)\ln {R\over r_c} 
\nonumber \\
-2\pi \hbar n \bm t \cdot[\bm R\times \tilde{\bm v}],
   \eem{Ven}
where $\bm t$ is the unit vector along the vortex axis, $\bm R$ is the  position vector for the vortex axis with its origin being either at a wall, or the position of the other vortex (antivortex), and
\bem
\tilde{\bm v} ={\hbar\over m} [ (N_\Phi - \sin \theta_\infty   N _\varphi )\bm K_\Phi
\nonumber \\
+ (N_\varphi - \sin \theta_\infty   N _\Phi)\bm K]={m \Gamma \over h}  \bm v+{N _\varphi \hbar \cos^2\theta_\infty \over m}\bm K 
    \eem{}
is the effective velocity.

The vortex energy $E_v$ has a maximum at
\be
R=\frac{\hbar (N_\Phi^2 -2\sin \theta_\infty N_\Phi N_\varphi + N_\varphi ^2)}{2 m \tilde v },
   \ee{}
and the energy at the maximum is a barrier preventing phase slips. The barrier vanishes if $R$ becomes of the order of the vortex core radius. Thus, phase slips are suppressed by energetic barriers as far as 
\be
\tilde v< \frac{\hbar (N_\Phi^2 -2\sin \theta_\infty N_\Phi N_\varphi + N_\varphi ^2)}{ m r_c }.
    \ee{vEff}
In single-component scalar superfluids the effective velocity $\tilde{\bm v}$ coincides with the superfluid velocity ${\bm v}$. When the latter vanishes the barrier is infinite, and phase slips are impossible. This reflects the trivial fact that the currentless state is the ground state, and phase slips cannot decrease its energy. In our case of a multicomponent superfluid phase slips also cannot decrease the energy if the effective velocity $\tilde{\bm v}$ vanishes.  But the latter vanishes not only in the currentless ground state but also in states with nonzero gradients satisfying the condition $\tilde{\bm v}=0$. Thus, the only one type of vortices  with fixed  ratio of two winding numbers is not sufficient for complete decay of  a current state with arbitrary values of two phase gradients. Complete relaxation to the ground state requires at least  two types of vortices with different ratios of winding numbers. 

For the vortex (0,1) with the core radius \eq{core01} at weak spin polarization $\theta_\infty\ll 1$ the inequality \eq{vEff} becomes
\be
K< \frac{1 }{  r_c}=\frac{\theta_\infty}{  \xi_0}.
    \ee{}
This condition imposes the same restriction on stability of the current state  as the Landau criterion \eq{gapL}  on stability of the gapped mode at small $\theta_\infty$.  Thus, instability of the gapped mode is a precursor of phase slips with this type of vortices. Phase slips only with vortices $(0,\pm 1) $  cannot result in complete relaxation of arbitrary current states to the ground state. A final state after  these phase slips is a state with zero effective velocity $\tilde{\bm v}$. The latter states lie on the dotted line in Fig.~\ref{fig2}(a).  

For the vortex  $\left({1\over 2},{1\over 2}\right)$ close to saturated spin polarization ($\theta_\infty \sim \pi/2$) with the core radius \eq{coreF} 
the inequality \eq{vEff} yields the inequality 
\be
{m v\over \hbar}+2 K < {1\over  r_c } ={\tilde \theta_\infty\over  \xi_0}, 
    \ee{}
which at small $\tilde \theta_\infty$ agrees with the Landau criterion \eq{GoldL} for the gapless mode. Thus, instability of this mode is a precursor of phase slips with ferromagnetic vortices  $\left({1\over 2},{1\over 2}\right)$. These phase slips result in relaxation of an initial arbitrary current state to current states with $\tilde{\bm v}=0$, which lie on the dotted line in  Fig.~\ref{fig2}(b). 

At complete saturation of spin polarization ($\theta_\infty =\pi/2$) no stable spin current is possible. The vortex $\left({1\over 2},{1\over 2}\right)$ has zero energy and zero velocity circulation and is irrelevant.  But mass persistent currents are possible as far as they are stable with respect to phase slips with vortices $\left(-{1\over 2},{1\over 2}\right)$. The core radius of this vortex is of the order of the coherence length $\xi_0$, and the condition of stability with respect to phase slips is similar to that obtained from the Landau criterion.

In our analysis the criterion of instability  was disappearance of energetic barriers suppressing the decay of mass or spin supercurrents. The critical velocities (gradients) are inversely proportional to the core radius of vortices participating in phase slips. In reality phase slips may  occur 
even in the presence of barriers due to thermal activation or quantum tunneling, although their probability is low. At low velocities (gradients) the logarithmic factor in the  vortex energy \eq{Ven} is very large and weakly depends on the value of the core radius $r_c$.  Then the chance of the vortex to participate in the phase slip is determined mostly by the pre-logarithmic factor in \eq{Ven}. The vortex with the smallest pre-factor is the most probable actor in the phase slip. This makes vortices with smaller circulations better candidates for phase slips. In particular, in the ferromagnetic spin-1 BEC at magnetic fields sufficient for spin orientation along magnetic fields  the half-integer vortex   $\left(-{1\over 2},{1\over 2}\right)$ with single-quantum circulation $h/m$ is a more probable actor in phase slips rather than the Anderson--Toulouse integer vortex   (-1,1)  with double-quantum velocity circulation $2h/m$ despite the latter has a larger core radius.

\section{Conclusions}

The hydrodynamics of the antiferromagnetic spin-1 BEC was derived from the Gross--Pitaevskii theory showing its analogy with the LLG theory of bipartite solid antiferromagnets. In the hydrodynamics of  spin-1 BEC two subspins with the absolute value $\hbar/2$ play the role of two sublattice spins in the antiferromagnetic insulators.  Thus, the Gross--Pitaevskii theory is a simple microscopic model justifying the LLG theory for antiferromagnets, while the microscopical derivation of the LLG theory for localized spins is problematic \cite{LLstPh2}.

The developed hydrodynamical theory was used for investigation of spin and mass supercurrents in the spin-1 BEC.
As well as in the ferromagnetic state, in the antiferromagnetic state of the spin-1 BEC the Landau criterion shows that stability of both mass and spin supercurrents is determined by the softest mode  which is one of spin-wave modes if the sound velocity exceeds essentially the spin wave velocity. Then one can check the Landau criterion only for spin wave modes assuming  that the superfluid is incompressible. In the antiferromagnetic state of the spin-1 BEC there are two spin-wave modes, one is a gapless Goldstone mode, another has a gap.  According to the Landau criterion, the gapped mode becomes unstable earlier than the gapless mode at small canting angles (weak spin polarization). In the opposite limit of the canting angle close to $\pi/2$ (two subspins are nearly parallel) the gapless mode becomes unstable earlier than the gapped one. 

The Landau instability is a precursor of the fast decay of supercurrents via phase slips. The paper analyzed what sort of vortices can participate in phase slips. These are nonsingular vortices with skyrmion cores without density suppression inside cores. The vortices are called bicirculation vortices because they are determined by two topological charges. The charges are winding numbers for circulations of two angle variables around the vortex axis. At the same time, the superfluid velocity circulation is not a topological charge because the superfluid velocity is not curl-free. 
The winding numbers of two angle variables can be half-integer. A particular case of half-integer vortices is a half-quantum vortex with the superfluid velocity circulation $h/2m$. However, in general one can tune continuously the velocity circulation quantum of a vortex between 0 and $h/2m$. This must have important consequences for properties of the spinor BEC, especially at its rotation.

\end{document}